**Singular Behavior of the Laplace Operator in Polar Spherical Coordinates and Some of Its Consequences for the Radial Wave Function at the Origin of Coordinates**

**Anzor A. Khelashvili[1,2] and Teimuraz P. Nadareishvili[1,3]**

[1]Institute of High Energy Physics, Iv. Javakhishvili Tbilisi State University, University Str. 9,

0109, Tbilisi, Georgia

[2] St. Andrea the First-called Georgian University of Patriarchy of Georgia, Chavchavadze Ave.

53a, 0162, Tbilisi, Georgia

E-mail: anzorkhelashvili@homail.com

[3] Iv. Javakhishvili Tbilisi State University, Faculty of Exact and Natural Sciences, Chavchavadze

Ave. 3, 0179, Tbilisi, Georgia

E-mail: teimuraz.nadareishvili@tsu.ge

**Abstract:** Singular behavior of the Laplace operator in spherical coordinates is investigated. It is shown that in course of transition to the reduced radial wave function in the Schrodinger equation there appears additional term consisting the Dirac delta function, which was unnoted during the full history of physics and mathematics. The possibility of avoiding this contribution from the reduced radial equation is discussed. It is demonstrated that for this aim the necessary and sufficient condition is requirement the fast enough falling of the wave function at the origin. The result does not depend on character of potential – is it regular or singular. The various manifestations and consequences of this observation are considered as well. The cornerstone in our approach is the natural requirement that the solution of the radial equation at the same time must obey to the full equation.

## 1. INTRODUCTION

The aim of this paper is to survey the singular behavior of the Laplacian in spherical coordinates. Laplacian is encountered almost in all disciplines of Theoretical physics as well as in mathematical physics. In this article our attention is paid mostly to the Schrodinger equation, which in the Cartesian coordinates has a form (in units $\hbar = c = 1$)

$$\left[ -\frac{1}{2m}\Delta + V(r) \right] \psi(\mathbf{r}) = E\psi(\mathbf{r}); \tag{1}$$

where

$$\Delta \equiv \nabla \cdot \nabla = \frac{\partial^2}{\partial x^2} + \frac{\partial^2}{\partial y^2} + \frac{\partial^2}{\partial z^2} \tag{2}$$

is a Laplacian.

In spherical coordinates the variables are separated and the total wave function is represented as

$$\psi(\mathbf{r}) = R(r)Y_l^m(\theta,\varphi) = \frac{u(r)}{r}Y_l^m(\theta,\varphi) \tag{3}$$

The Laplacian is also rewritten in terms of these coordinates and after the substitution of Eq. (3) into the Eq. (1) we derive the radial equations

$$-\frac{1}{2m}\left[\frac{d^2}{dr^2} + \frac{2}{r}\frac{d}{dr}\right]R(r) + \frac{l(l+1)}{2mr^2}R(r) + V(r)R(r) = ER(r) \tag{4}$$

or

$$\left[ -\frac{1}{2m}\frac{d^2}{dr^2} + \frac{l(l+1)}{2mr^2} + V(r) \right] u(r) = Eu(r) \tag{5}$$

All of this is well known from the classical textbooks on quantum mechanics, electrodynamics and etc. We display them here for further practical purposes. It will be shown below that the status of the Eq. (5) is problematic.

From both mathematical and physical points of view it is very important that the solutions of radial equations were compatible with the full Schrodinger equation (1).This is verbally mentioned in books, not only earlier [1,2] but also in the modern ones [3]. For example, Dirac [1] wrote: *"Our equations ... strictly speaking, are not correct, but the error is restricted by only one point $r = 0$. It is necessary perform a special investigation of solutions of wave equations, that are derived by using the polar coordinates, to be convince are they valid in the point $r = 0$ (p.161)"*

We are sure that mathematicians new about this problem (singularity of the Laplacian) for a long time, but character of singularity never been specified. It was always underlined in mathematics that $r > 0$ strictly, but $r = 0$ is not somehow prominent point for the 3-dimensional equation. Therefore refinement of the behavior of the radial wave function at that point has a basic meaning by our opinion.

The first papers [4-7] on this problem appeared recently almost in parallel.

Because of relative novelty of this subject below we will take some attention to its substantiation.

To complete the picture we first discuss briefly the essence of this problem and then some of its application will be considered.

In the teaching books and scientific articles two methods were applied in the transition from Eq. (4) to Eq. (5):

1. Substitution

$$R(r) = \frac{u(r)}{r} \qquad (6)$$

into the Eq. (4) or

2. Replacement of the differential expression in the parenthesis of Eq. (4) as [8-10]*

$$\left[\frac{d^2}{dr^2} + \frac{2}{r}\frac{d}{dr}\right] \rightarrow \frac{1}{r}\frac{d^2}{dr^2}(r.) \qquad (7)$$

We demonstrate below that in both cases the mistakes were made.

Because all the principal information is concentrated in the Laplace operator, we begin by consideration the classical Laplace equation in the vacuum (electrostatic equation)

## 2. THE LAPLACE EQUATION

Let us consider the Laplace equation in vacuum

$$\nabla^2 \varphi(\vec{r}) = 0 \qquad (8)$$

which in Cartesian coordinates has the form

$$\nabla^2 \varphi(\mathbf{r}) = \left(\frac{\partial^2}{\partial x^2} + \frac{\partial^2}{\partial y^2} + \frac{\partial^2}{\partial z^2}\right) \varphi(x, y, z) = 0 \qquad (9)$$

This equation may be solved simply by separation of variables. The solution has the form [10]

$$\varphi(x, y, z) = e^{\pm i\alpha x} e^{\pm i\beta y} e^{\pm\sqrt{\alpha^2 + \beta^2}\, z} \qquad (10)$$

Clearly the solution is regular everywhere and at the origin is constant

$$\varphi(0) = const \qquad (11)$$

There are another forms of solution of Eq.(9) depending on alternate ways of separation, but all of them give the constant values at the origin.

Now, let us find the spherically symmetric solution. The corresponding equation is written as [8]

$$\left(\frac{d^2}{dr^2} + \frac{2}{r}\frac{d}{dr}\right)\varphi(r) = 0 \qquad (12)$$

Certainly, it was possible passing to spherical coordinates in Eq. (9), substituting (3) and taking zero angular momentum. We'll arrive again to the Eq. (12).

The operator in parenthesis of Eq. (12) often is rewritten ([8], Ch.20, [9] etc.) according to (7) and subsequently, equation (12) takes the form

$$\frac{1}{r}\frac{d^2}{dr^2}(r\varphi) = 0 \qquad (13)$$

the solution of which is

$$u(r) \equiv r\varphi = ar + b \qquad (14)$$

But, determining from here the function

---

* In the fundamental book of J.D.Jackson [10] this relation is even exhibited on the cover-page in the list of the most fundamental forms!

$$\varphi = a + \frac{b}{r} \tag{15}$$

does not obey to Eq. (12), because

$$\left(\frac{d^2}{dr^2} + \frac{2}{r}\frac{d}{dr}\right)\left(\frac{1}{r}\right) = -4\pi\delta^{(3)}(\mathbf{r}) \tag{16}$$

i.e. the function (15) is the solution everywhere except the origin of coordinates. It does not satisfy the boundary value (11) as well.

What happens? *It seems that we made an illegal action somewhere* (see, Feynman [8]).

It is possible to consider this problem by another way also, namely, following to the substitution (6), take

$$\varphi(r) = \frac{u(r)}{r} \tag{17}$$

in order to remove the first derivative term from the Eq. (12). Then we obtain

$$\frac{1}{r}\left(\frac{d^2}{dr^2} + \frac{2}{r}\frac{d}{dr}\right)u(r) + u(r)\left(\frac{d^2}{dr^2} + \frac{2}{r}\frac{d}{dr}\right)\left(\frac{1}{r}\right) + 2\frac{du}{dr}\frac{d}{dr}\left(\frac{1}{r}\right) = 0 \tag{18}$$

The last term cancels the first derivative term in the first parenthesis and there remains

$$\frac{1}{r}\frac{d^2u}{dr^2} + u\left(\frac{d^2}{dr^2} + \frac{2}{r}\frac{d}{dr}\right)\left(\frac{1}{r}\right) = 0 \tag{19}$$

but, according to Eq. (16), instead of Eq. (13), it follows

$$\frac{1}{r}\frac{d^2u}{dr^2} - 4\pi\delta^{(3)}(\mathbf{r})u(r) = 0 \tag{20}$$

The appearance of the delta function here is unexpected. Comparing this one with Eq. (13) we conclude that the representation of the Laplace operator in the form (7) is not valid *everywhere*. The correct form is [5, 7]

$$\frac{d^2}{dr^2} + \frac{2}{r}\frac{d}{dr} = \frac{1}{r}\frac{d^2}{dr^2}(r\cdot) - 4\pi\delta^{(3)}(\mathbf{r})r\cdot \tag{21}$$

This expression defines the form of the Laplasian precisely everywhere including the origin of coordinates.

It is evident, that after substitutions

$$\left(\frac{d^2}{dr^2} + \frac{2}{r}\frac{d}{dr}\right)\varphi \Rightarrow \frac{1}{r}\frac{d^2}{dr^2}(r\varphi) \quad \text{and} \quad u = r\varphi, \tag{22}$$

the solution $\varphi = u/r$, obtained from the equation (13), never satisfies to the initial equation (12) *everywhere*.

By unknown for us reasons this simple fact stayed unnoted till now and in all papers as well as in all books the expression (7) was used. As we made clear up above, in this case the obtained solution (15) looks like if there is a point source at the origin. However it is not so – mathematic reason is that in spherical coordinates the point $r = 0$ is absent. The Jacobian of transformation to spherical coordinates has a form $J = r^2\sin\theta$ and is singular at points $r = 0$ and $\theta = n\pi \quad (n = 0,1,2,...)$.

Singularity in angles is eliminated by requirements of continuity and uniqueness, which lead to spherical harmonics $Y_l^m(\theta,\varphi)$. As regards of the radial variable r, there is no such restriction for it. Therefore if we want to derive the solution valid everywhere, we are forced to include the delta function into the consideration.

It must be noted that the appearance of the delta function in the Laplace equation was discussed also in article [6], where the difference between spaces $R^n$ and $R^n/\{0\}$ is studied from the positions of distribution theory.

The question is: *how to formulate the problem in such a way that to remain all results derived earlier for the central potentials with the aid of traditional reduced radial equation (5) containing the second derivative only?* One of the reasonable ways is the following: Because in spherical coordinates $\delta^{(3)}(\vec{r}) = \dfrac{\delta(r)}{4\pi r^2}$ [11], the Eq. (20) can be reduced to

$$r\frac{d^2 u}{dr^2} - \delta(r)u(r) = 0 \tag{23}$$

or

$$r\frac{d^2 u}{dr^2} - u(0)\delta(r) = 0 \tag{24}$$

Let us require that the additional term does not present i.e.
$$u(0) = 0 \tag{25}$$
Moreover the delta function be "overcome" if at least
$$\lim_{r\to 0} u(r) \approx r \tag{26}$$

Then, owing to the relation $r\delta(r) = 0$, the extra term falls out and the standard equation (13) follows. Let us look first what the condition (25) gives in above considered solution (14). Requiring (25), it follows $b = 0$, i.e. $u = ar$ and $\varphi(r) = a = const$. Hence we obtain the correct, consisting with the full equation (8) value (11). It is consisting also with the real physical picture.

Therefore in the reduced radial equation (5) we must consider only such class of solutions, which vanish at the origin. The other entire boundary conditions loss the physical meaning and have only mathematical interest. It is precisely the main result of this section – the equation (5) gives the consistent with the primary equation in Cartesian coordinate's solution only if the restriction (25) is satisfied. Appearance of this condition is purely geometrical (not a dynamical) artefact. In short words, the Eq. (5) and the condition (25) appear simultaneously.

## 3. THE RADIAL SCHRODINGER EQUATION AND $u(0)$

As an example let us consider the radial Schrodinger equation (4)
After the substitution (6), according to above mentioned about the Laplace operator, we obtain the following form of this equation

$$\frac{1}{r}\left[\frac{d^2 u(r)}{dr^2} - \frac{l(l+1)}{r^2}u(r)\right] - \frac{\delta(r)}{r^2}u(r) + 2m[E - V(r)]\frac{u(r)}{r} = 0 \tag{27}$$

To single out the true singularity let us multiply this equation on $r^2$ and integrate by $dr$ in a sphere of small radius $a$. We derive

$$\int_0^a r\frac{d^2 u(r)}{dr^2}dr - l(l+1)\int_0^a \frac{u(r)}{r}dr - u(0) + \int_0^a (2mE - V(r))u(r)r\,dr = 0 \tag{28}$$

From here we determine

$$u(0) = \int_0^a r \frac{d^2u(r)}{dr^2}dr - l(l+1)\int_0^a \frac{u(r)}{r}dr + \int_0^a (2mE - V(r))u(r)r dr \qquad (29)$$

Because of smallness of $a$ substitute here the asymptotic form of wave function at the origin

$$u(r) \underset{r \to 0}{\approx} r^s \qquad (30)$$

and the potential as

$$V(r) \underset{r \to 0}{\approx} \frac{g}{r^n}; \qquad n > 0 \qquad (31)$$

Then the integration in Eq. (29) may be easily performed and we obtain

$$u(0) = \left[ \frac{s(s-1) - l(l+1)}{s} r^s + \frac{2mE}{s+2} r^{s+2} - \frac{2mg}{s+2-n} r^{s+2-n} \right]_0^a \qquad (32)$$

We must remove the extra delta term from Eq. (27), because otherwise we do not get the usual form of radial equation (5).

If we retain $u(0)$ in the Eq. (28) then there are 3 possible values for it: $u(0) = 0$, $u(0)$ is finite and $u(0) = \infty$. Note that all the enumerated values do not contradict to normalization condition near the origin $\int_0^a u^2 dr < \infty$, but not all of them are useful.

The first value is preferable among them, because in opposite cases - finite $u(0)$ will give $R \approx \frac{const}{r}$ at the origin and in Eq. (27) the delta function reappears again. Therefore this solution will not obey to full Schrodinger equation. The last value, $u(0) = \infty$ of course is unacceptable, because to have an infinite number in equation is a senseless.

There remains only one reasonable value, Eq. (25). Moreover this restriction takes place in spite of the potential is regular or singular. Singularity of the potential effects only on the law of turning of $u(r)$ to zero. This follows from the relation (32) as all the exponents here must be positive. We'll have therefore

$$s > 0, \quad s+2 > 0, \quad s+2-n > 0$$

It follows from the last inequality that when the index of singularity of potential n increases, the index of wave function behavior $s$ must also increase. Moreover we must have $s \geq 1$ in order the wave function at the origin "overcomes" the delta function in the term $u(r)\delta(r)$. Therefore there remains the final allowed inequalities

$$s \geq 1, \qquad s+2-n > 0 \qquad (33)$$

If in addition we require that this production be a distribution, it becomes necessary that $u(r)$ be an infinitely smooth function [12, 13], i.e. in Eq. (30) we must have $s \geq 1$ and the index $s$ is an integer number.

Thus the wave function must be sufficiently regular one at the origin. This fact may have far reaching consequences.

## 4. SOME APPLICATIONS

The first question, that appears here, is the following: under what conditions can we maintain the standard form of reduced wave equation?

Basing on the previous considerations we suppose that the equation in the standard form (5) takes place and clarify for which potentials it happens, i.e. when we can satisfy the restriction (25)?

### 4A) Regular potentials

Let us consider first the regular potentials

$$\lim_{r\to 0} r^2 V(r) = 0 \tag{34}$$

Then in the Schrodinger equation (5) the leading asymptotic at the origin is determined by a centrifugal term and the characteristic equation takes the form $s(s-1) = l(l+1)$. So

$$u \underset{r\to 0}{\sim} c_1 r^{l+1} + c_2 r^{-l}; \; l = 0,1,2... \tag{35}$$

We must retain only the first solution because now $s = l+1 \geq 1$ and the derived representation is satisfied (s is an integer number!). At the same time the second solution with $s = -l$ must be ignored even for $l = 0$ [14]. The second solution does not satisfy to the 3-dimensional Schrodinger equation (1), as after its substitution the Laplacian produces $l$-fold derivatives of delta function [14].

Resuming above saying we conclude that in case of regular potentials (34) the radial equation (5) remains, because in this case the all requirements are realized and consequently, the results obtained earlier by this equation remain valid without any changes!

### 4B) Weakly singular transitive potentials

Let us now consider potentials that are intermediate between singular and regular ones, so-called weakly-singular potentials of the form

$$\lim_{r\to 0} r^2 V(r) = -V_0 = const \tag{36}$$

Here $V_0 > 0$ corresponds to the attraction, while $V_0 < 0$ - to repulsion. Now the behavior of $u(r)$ at the origin is

$$u \underset{r\to 0}{\sim} d_1 r^{\frac{1}{2}+P} + d_2 r^{\frac{1}{2}-P}; \quad P = \sqrt{\left(l+\frac{1}{2}\right)^2 - 2mV_0} > 0 \tag{37}$$

In order that the usual equation (5) will still remain, according to Eq.(33) we must have $(s \geq 1)$, i.e. $P \geq \frac{1}{2}$ for all $l$, including $l = 0$, and at the same time, according to requirement of the distribution theory $1/2 + P = N$; $N = 1,2,3...$ So it results a "strange quantization" of $V_0$, which is also senseless. It follows that in this case there are no solutions except for "quantized" $V_0$.

We see that the second solution in Eq. (37) must be discarded. Note that in scientific literary there is no definite viewpoint concerning to this. (see, e.g. book by R. Newton [15] and various modern articles [16,17]). Therefore the above mentioned derivation is a first correct one.

### 4C) The problem of a Self-adjoint extention (SAE)

Last decades the problem of self-adjoint extension (SAE) of radial Hamiltonian

$$H_r \equiv -\frac{d^2}{dr^2} + \frac{l(l+1)}{r^2} + 2mV(r) \tag{38}$$

was often considered in cases of singular potentials, like above one. In this problem the essential role plays the behavior of radial wave functions $u(r)$ at the origin of coordinates. For example, the condition of self-adjointicity of Hamiltonian (38) has a form [18]

$$\int_0^\infty u_1 \hat{H}_r u_2 dr - \int_0^\infty u_2 \hat{H}_r u_1 dr = \frac{1}{2} \lim_{r \to 0} [u_2(r) u_1'(r) - u_1(r) u_2'(r)] = 0 \tag{39}$$

where $u_{1,2}(r) \equiv r R_{1,2}(r)$ are two linearly independent solutions of the reduced radial equation (5) corresponding to different eigenvalues of the Hamiltonian (38). There were considered various boundary conditions such as ones of Dirichlet, Neuman, and the most general condition of Robin [19]. *While as we made clear above only the Dirichlet condition (25) is right*

In most articles in course of discussion of SAE procedure with the Hamiltonian (38) authors take attention only on square integrability of the wave function [20]. But it is not sufficient in all cases. Still W.Pauli [21] noted that "the eigenfunction, for which $\lim_{r \to 0}(rR) \neq 0$, is not permissible (even if $\int_0^\infty R^* R r^2 dr$ exists for such functions)". The same is confirmed in the more modern books (for example, in [3, pf 52] the author says: "It can be shown that the condition u(0)=0 follows from the requirement that the solution of the Schrodinger equation in spherical coordinates must be also solution when the equation is written in Cartesian coordinates"). But unfortunately, this thesis is not shown regularly in this book, especially for singular potentials.

If we impose the boundary condition with the indices $s \geq 1$ we must restrict ourselves only by the first (regular) solution, i.e. $d_2 = 0$ (See, Eq. (37)). Then the radial Hamiltonian (38) becomes a self-adjoint one automatically and the SAE is not needed. As for the first solution the condition $P \geq 1/2$ is achieved only if

$$l(l+1) > 2mV_0 \tag{40}$$

i.e. for $l = 0$ only $V_0 < 0$ is permissible and as regards of other admissible values, from the condition $1/2 + P = N$ it follows a strange ''quantization'' of $V_0$

$$V_0 = \frac{(l+1/2)^2 - (N-1/2)^2}{2m}; \quad N = 1,2,3... \tag{41}$$

Hence even for such a simple singular potential (36) the equation (5) meets the serious physical difficulties.

We do not consider here the other, more singular potentials, because the general tendency is obvious. The Hamiltonian (38) by itself is always a self-adjoint on the regular solutions, satisfying to (25), as it follows from the condition (39) and restrictions (33) for any singular potentials. For all other boundary conditions the Hamiltonian (38) will not have bear a relation to physics, because this form of Hamiltonian emerges only together with condition (25).

We conclude that the reduced radial equation (5) may be applied for all regular potentials, nevertheless for singular potentials one must work only with the total radial equation (4) and, consequently, use the full radial Hamiltonian

$$H_R = -\frac{d^2}{dr^2} - \frac{2}{r}\frac{d}{dr} + \frac{l(l+1)}{r^2} + 2mV(r) \tag{42}$$

but search for regular solutions only. In the article [4] we have shown that from the finiteness of the differential probability $dW = |R(r)|^2 r^2 dr$ and the time independence of the norm it follows that $R(r)$ is less singular at the origin, than $1/r$ or

$$\lim_{r \to 0} rR = 0 \tag{43}$$

which, evidently is consistent to $u(0) = 0$.

Moreover in case of fulfillment this condition the radial equation (4) for full radial function $R(r)$ is equivalent to the Schrodinger equation (1). This equivalence takes place only on non-singular solutions. In other words, the Eq. (4) is equivalent to the 3-dimensional Schrodinger equation only for regular solutions. This was proved also in paper [6] in the framework of the distribution theory.

For demonstration of principal difference between the full and reduced radial Hamiltonians let us consider now the same problem in view of the full radial function $R(r)$. The condition (43) is the only boundary restriction for it, which is not so severe. Therefore there appears a possibility to retain the second solution as well in the case of singular potentials behaving like (36)

The following statement can be proved:

<u>Theorem.</u>The radial Schrodinger equation (4) except the standard (non-singular) solutions has also additional solutions for attractive potentials, like (36), when the following condition is satisfied

$$l(l+1) < 2mV_0 \tag{44}$$

. The proof of this theorem is straightforward.

Indeed, for the attractive potential (36) <u>at small distances</u> this equation reduces to

$$R'' + \frac{2}{r}R' - \frac{P^2 - 1/4}{r^2} R = 0 \tag{45}$$

where $P$ is defined by (37).

Therefore, Eq. (45) has following solution

$$\lim_{r \to 0} R = a_{st} r^{-1/2+P} + a_{add} r^{-1/2-P} \equiv R_{st} + R_{add} \tag{46}$$

So we have two regions for this parameter $P$. In the interval

$$0 < P < 1/2 \tag{47}$$

the second term $a_{add} r^{-1/2-P} = R_{add}$ must be also retained, because the boundary condition (43) is fulfilled for it. The potential like (36) was first considered by K.Case [22], but he ignored the second term in solution. As regards of a region $P \geq \frac{1}{2}$, only the first term $a_{st} r^{-1/2+P} = R_{st}$ must be retained.

From eqs. (37) and (47) it follows the condition (44) for existence of additional states. If we demand the reality of $P$ (otherwise ''falling'' to centre takes place [22,23,24]) the parameter $V_0$ would be restricted by condition

$$2mV_0 < l(l+1) + 1/4 \tag{48}$$

The last two inequalities restrict $2mV_0$ in the following interval

$$l(l+1) < 2mV_0 < l(l+1) + 1/4 \tag{49}$$

Intervals from the left and from the right sides have no crossing and therefore, if additional solution exists for fixed $V_0$ and for some $l$, then it is absent for another $l$.

Thus we see from (44) that in the $l=0$ state except the standard solutions there are additional solutions too for arbitrary small $V_0$, while for $l \neq 0$ the "strong" field is required in order to fulfil (44).

Because the additional solutions obey all physical requirements in the interval (47), one has to retain this solution as well and study its consequences.

For definiteness consider the potential

$$V = -\frac{V_0}{r^2}, \qquad V_0 > 0 \qquad (50)$$

When E=0 the solution of the full radial equation (4) has the form in whole space

$$R = A r^{-1/2+P} + B r^{-1/2-P} \qquad (51)$$

There is only one worthy case, namely $0 < P < 1/2$. We see that the wave function has a simple zero, determined by

$$r = r_0 = \left(-\frac{B}{A}\right)^{1/P} \qquad (52)$$

(It is evident from this relation that constants $A$ and $B$ must have opposite signs in order for $r_0$ to be real number). Hence, the wave function has only one node and according to well-known theorem (*the number of bound states coincides with the number of nodes of radial wave function $R(r)$ in $E=0$ state* [2]), we have exactly one bound state.

This result differs from that considered in any textbooks on quantum mechanics.

We can give very simple physical picture of how the additional solutions arise. For this purpose, let us rewrite the Schrodinger equation near the origin for attractive potential (36) in the form

$$R'' + \frac{2}{r} R' + 2m[E - V_{ac}(r)]R = 0 \qquad (53)$$

where

$$V_{ac} = \frac{P^2 - 1/4}{2mr^2} \qquad (54)$$

Consider the following possible cases:

i). If $P > 1/2$, then $V_{ac} > 0$ and it is repulsive centrifugal potential and as we saw, one has no additional solutions.

ii). If $0 < P < 1/2$, then $V_{ac} < 0$. Therefore, it becomes attractive and is called as quantum anti-centrifugal potential [25]. This potential has $R_{add}$ states, because the condition (43) is fulfilled in this case.

iii). If $P^2 < 0$, then $V_{ac}$ becomes strongly attractive and one has "falling to the center".

Therefore, the sign of the potential $V_{ac}$ determines whether we need additional solutions or not.

### 4D) SAE procedure for full Radial Hamiltonian in "pragmatic" approach

Considering some consequences from the point of above mentioned results, let us first of all remember some issues of SAE procedure.

If for any functions $u$ and $\upsilon$, given operator $\hat{A}$ satisfies to the condition

$$\langle \upsilon | \hat{A} u \rangle = \langle \hat{A} \upsilon | u \rangle \qquad (55)$$

then this operator is called hermitian (or symmetric). For self-adjointness it is required in addition that the domains of functions of operators $\hat{A}$ and $\hat{A}^+$ would be equal. As a rule, the domain of the $\hat{A}^+$ is wider and it becomes necessary to make a self-adjoint extension of the operator $\hat{A}$.

There exists a well-known powerful mathematical apparatus for this purpose [26, 27].

It may happen that the operator is hermitian, but its self-adjoint extension is impossible, i.e. hermiticity is the *necessary, but not sufficient condition* for self-adjointness. Good example is the operator of the radial momentum $p_r$ which is hermitian on functions that satisfy the condition (43), but its extension to self-adjoint one is impossible (see, L.D.Faddeev's remark in the A.Messiah's book – Russian translation, footnote in p.336 [28]).

Our subject of interest is the radial Hamiltonian (42) and, consequently, the equation (4)

It is easy to see that for any two eigenfunctions $R_1$ and $R_2$ corresponding to the levels $E_1$ and $E_2$ of the radial Hamiltonian $\hat{H}_R$, the condition (55) takes the following form

$$\int_0^\infty R_1 \hat{H}_R R_2 r^2 dr - \int_0^\infty R_2 \hat{H}_R R_1 r^2 dr = 2m(E_2 - E_1)\int_0^\infty R_1 R_2 r^2 dr \qquad (56)$$

It follows that a self adjoint condition is proportional to the orthogonality integral, therefore these two conditions are mutually dependent. Because the self-adjoint operator has orthogonal eigenfunctions, requirement of orthogonality automatically provides self-adjointness of $H_R$, which means that this way provides realization of SAE procedure. It is an essence of so-called "pragmatic approach" [29], which is much simpler and gets the same results as the strong mathematical full SAE procedure, provided the fundamental condition (43) is not violated. Moreover this method is physically more transparent. Just this method had been used by Case in his well-known paper [22], but he did consider only the regular solution.

Notice that all above considerations are true only for the radial Hamiltonian operator $\hat{H}_R$, because for other operators proportionality like (56) does not arise.

## 4E) Explicit solution of the Schrodinger equation for the inverse squared potential

It was thought that potential (50) has no levels out of region of "falling to the center" (See e.g. [22,23]), but in [16,20,30] single level was found by complete SAE procedure, while the boundary condition and the range of parameter, like P are questionable there. Here we'll show explicitly that this potential has exactly single level, which depends on the SAE parameter τ.

Let us now study in which cases the right-hand sides of (56) is vanishing. We must distinguish regular and transitive potentials. As we are interested of bound states we suppose that the full radial function decreases suficiently fast at infinity. So, the behavior at the origin is relevant for our aims.

In case of regular potentials (34), as was mentioned above, we retain only first, regular (or standard) solution at the origin

$$R_{st} \underset{r \to 0}{\sim} a_{st} r^{l+1} \qquad (57)$$

Calculating the r.-h.-side of (56) by this function, we get zero. Therefore for regular potentials the radial Hamiltonian $\hat{H}_R$ is self-adjoint on regular solutions and it does not need SAE.

Contrary to this case, for transitive attractive (36) potential one has to retain the additional solution $R_{add} \underset{r \to 0}{\sim} r^{-1/2-P}$ as well, because there are no reasons to neglect it. Now for both solutions, the r.-h.-sides of (56) are not zero in general. Indeed they equal to

$$m(E_1 - E_2) \int_0^\infty R_2 R_1 r^2 dr = P\left(a_1^{st} a_2^{add} - a_2^{st} a_1^{add}\right) \qquad (58)$$

*Remark*. The case $P = 0$ must be considered separately, when the general solution of (4) behaves as

$$\lim_{r \to 0} R = a_{st} r^{-\frac{1}{2}} + a_{add} r^{-\frac{1}{2}} \ln r = R_{st} + R_{add} \qquad (59)$$

So, instead of (58) one obtains

$$m(E_1 - E_2) \int_0^\infty R_2 R_1 r^2 dr = -\frac{1}{2}\left(a_1^{st} a_2^{add} - a_2^{st} a_1^{add}\right) \qquad (60)$$

Thus retaining additional solution causes the breakdown of orthogonality condition and consequently, $\hat{H}_R$ is no more a self-adjoint operator.

It is natural to ask – how to fulfil the orthogonality condition? It is clear, that in both $P \neq 0$ and $P = 0$ cases one must require

$$a_1^{st} a_2^{add} - a_2^{st} a_1^{add} = 0 \qquad (61)$$

or equivalently

$$\frac{a_{1\,add}}{a_{1\,st}} = \frac{a_{2\,add}}{a_{2\,st}} \qquad (62)$$

In this case the radial Hamiltonian $\hat{H}_R$ becomes a self-adjoint operator. This generalizes the Case result [22], who considered only standard solution.

So it is necessary to introduce so called SAE parameter, which in our case may be defined as

$$\tau \equiv \frac{a_{add}}{a_{st}} \qquad (63)$$

$\tau$ parameter is the same for all levels (for fixed orbital $l$ momentum) and is real for bound states.

Now let us return to the solution of the Schrodinger equation for potential (50)

$$\frac{d^2 R}{dr^2} + \frac{2}{r}\frac{dR}{dr} + \left(-k^2 - \frac{P^2 - 1/4}{r^2}\right) R = 0 \qquad (64)$$

where P is given by (37) and

$$k^2 = -2mE > 0; \quad (E < 0) \qquad (65)$$

One can reduce Eq. (64) to the equation for modified Bessel functions by substitutions

$$R(r) = \frac{f(r)}{\sqrt{r}}; \quad x = kr \qquad (66)$$

leading to the following equation

$$x^2 \frac{d^2 f(x)}{dx^2} + x \frac{df(x)}{dx} - \left(x^2 + P^2\right) f(x) = 0 \qquad (67)$$

This equation has 3 pairs of independent solutions: $I_P(kr)$ and $I_{-P}(kr)$, $I_P(kr)$ and $e^{i\pi P} K_P(kr)$, $I_{-P}(kr)$ and $e^{i\pi P} K_P(kr)$, where $I_P(kr)$ and $K_P(kr)$ are Bessel and MacDonald modified functions, respectively [31].

Careful analysis gives that the relevant pair is the first one only, i.e. the pair $I_P(kr)$ and $I_{-P}(kr)$:

So, the general solution of (64) is

$$R = r^{-\frac{1}{2}}[AI_P(kr) + BI_{-P}(kr)] \qquad (68)$$

Consider the behaviour of this solution at small and large distances:

a) Small distances

In this case (see, [31])

$$I_P(z) \underset{z \to 0}{\approx} \left(\frac{z}{2}\right)^P \frac{1}{\Gamma(P+1)} \qquad (69)$$

Then it follows from (68) and (69) that

$$\lim_{r \to 0} R(r) \approx r^{-\frac{1}{2}}\left[A\left(\frac{k}{2}\right)^P \frac{r^P}{\Gamma(P+1)} + B\left(\frac{k}{2}\right)^{-P} \frac{r^{-P}}{\Gamma(1-P)}\right] \qquad (70)$$

From (46), (62), (70) and the definition (63) we obtain

$$\tau = \frac{B}{A} 2^{2P} k^{-2P} \qquad (71)$$

At large distances, we have [31]

$$I_P(z) \underset{z \to \infty}{\approx} \frac{e^z}{\sqrt{2\pi z}} \qquad (72)$$

and

$$R(r) \underset{r \to \infty}{\approx} \frac{1}{\sqrt{2\pi}} \{A + B\} e^{kr} \qquad (73)$$

Therefore, requiring vanishing of $R(r)$ at infinity, we have to take

$$B = -A \qquad (74)$$

and from (71), (74) and (65) we obtain one real level (for fixed orbital $l$ momentum, satisfying (44)),

$$E = -\frac{2}{m}\left[-\frac{1}{\tau}\right]^{\frac{1}{P}}; \quad 0 < P < 1/2 \qquad (75)$$

Eq. (75) is a new expression derived as a consequence of orthogonality condition in the framework of "pragmatic" approach.

Reality of energy in (75) restricts $\tau$ parameter to be negative $\tau < 0$. In general $\tau$ is a free parameter, but some physical requirements may restrict its magnitude. Note that this level is absent in standard quantum mechanics $(\tau = 0)$ - it appears when one performs SAE procedure.

To obtain corresponding wave function, take into account a well-known relation [31]

$$K_P(z) = \frac{\pi}{2 \sin P\pi}[I_{-P}(z) - I_P(z)] \qquad (76)$$

Then the wave function corresponding to the level (75) be

$$R = -A\frac{2}{\pi} r^{-\frac{1}{2}} \sin P\pi \cdot K_P(kr) \qquad (77)$$

Because of exponential damping

$$K_P(z) \underset{z \to \infty}{\approx} \sqrt{\frac{\pi}{2z}} e^{-z} \qquad (78)$$

the function (77) corresponds to the bound state. It is also known that $K_P(z)$ function has no zeroes for real P ($0 < P < 1/2$) and therefore (75) corresponds to a single bound state. Moreover, wave function (77) satisfies the fundamental condition (43) for $0 < P < 1/2$.

Let us make some comments

a) In [20] it was noticed that single bound state may be observed experimentally in polar molecules. For example, $H_2S$ and $HCl$ atoms exhibit anomalous electron scattering [32,33], which can be explained only by electron capture. Indeed, for those molecules electron is moving in a point dipole field, and, in this case the problem is reduced to the Schrodinger equation with a potential (50). Thus, a level (75) obtained theoretically may be observed in those experiments.

b) It was commonly believed, that the potential

$$V = -\frac{V_0}{sh^2 \alpha r} \tag{79}$$

has no levels in region (47) (see, for example, problem 4.39 in [34]). In [34] by the arguments of well-known comparison theorem [26], which in this case looks like

$$-\frac{V_0}{sh^2 \alpha r} \geq -\frac{V_0}{\alpha^2 r^2} \tag{80}$$

it is concluded that the potential (79) cannot have a level in the area (47), because the potential (50) has no levels in this area. But, as we know, there is $\tau$ depended one level (75), therefore the levels for (79) are expected. Indeed, in [35] by using the Nikiforov-Uvarov method [36], it was shown that the potential (79) has infinite number of levels in the region (47).

## 5. OTHER APPLICATIONS

There are physically more realistic potentials, which differ from (50), but behave as $r^{-2}$ at the origin.

Famous examples are molecular potential (valence electron model), Coulomb potential in Klein- Gordon equation and etc.

### 5A) Valence electron model

Let us consider a molecular potential, having the following form

$$V = -\frac{V_0}{r^2} - \frac{\alpha}{r} \; ; \quad (V_0, \alpha > 0) \tag{81}$$

Because of a singular $r^{-2}$ like behavior at the origin one must consider equation for the $R(r)$ function, which in dimensionless variables takes form

$$\left(\frac{d^2}{d\rho^2} + \frac{2}{\rho}\frac{d}{d\rho} - \frac{P^2 - 1/4}{\rho^2} + \frac{\lambda}{\rho} - \frac{1}{4}\right)R = 0 \tag{82}$$

where

$$\rho = \sqrt{-8mE}\, r = ar; \quad \lambda = \frac{2m\alpha}{\sqrt{-8mE}} > 0, \quad E < 0 \tag{83}$$

and $P$ is again given by Eq.(37).

If we substitute

$$R = \rho^{-\frac{1}{2}+P} e^{-\frac{\rho}{2}} F(\rho), \tag{84}$$

the equation for confluent hypergeometric functions follows
$$\rho F'' + (2P + 1 - \rho)F' - (1/2 + P - \lambda)F = 0 \quad (85)$$

This equation has four independent solutions, two of which constitute a fundamental system of solutions [37]. They are (in notations of [37]):

$$y_1 = F(a,b;\rho)$$
$$y_2 = \rho^{1-b} F(1+a-b, 2-b; \rho)$$
$$y_5 = \Psi(a,b;\rho)$$
$$y_7 = e^\rho \Psi(b-a,b;-\rho)$$
(86)

where
$$a = 1/2 + P - \lambda, \qquad b = 1 + 2P \quad (87)$$

Only $y_1$ is considered in the scientific articles, as well as in all textbooks (see, e.g. [23,38]). Requiring $a = -n$ ($n = 0,1,2,...$) the standard levels follow. Other solutions ($y_2, y_5, y_7$) have singular behavior at the origin and usually they are not taken into account. But the singularity in case of attractive potentials like (36) has the form $r^{-\frac{1}{2}-P}$ and in the region $0 < P < 1/2$ other solutions must be considered as well. Therefore, the problem becomes more "rich".

Let us consider a pair $y_1$ and $y_2$. The general solution of (85) is

$$R = C_1 \rho^{-1/2+P} e^{-\frac{\rho}{2}} F(1/2+P-\lambda, 1+2P; \rho) + C_2 \rho^{-1/2-P} e^{-\frac{\rho}{2}} F(1/2-P-\lambda, 1-2P; \rho) \quad (88)$$

Considering Eq. (88) at the origin and accounting Eq. (63), we obtain the following expression for SAE $\tau$ parameter

$$\tau = \frac{C_2}{C_1} \frac{1}{(-8mE)^P} \quad (89)$$

On the other hand that, R must decrease at infinity. From well-known asymptotic properties of confluent hypergeometric function F, we find the following restriction

$$C_1 \frac{\Gamma(1+2P)}{\Gamma(1/2+P-\lambda)} + C_2 \frac{\Gamma(1-2P)}{\Gamma(1/2-P-\lambda)} = 0 \quad (90)$$

It gives an equation for eigenvalues in terms of $\tau$ parameter

$$\frac{\Gamma(1/2 - \lambda - P)}{\Gamma(1/2 - \lambda + P)} = -\tau(-8mE)^P \frac{\Gamma(1-2P)}{\Gamma(1+2P)} \quad (91)$$

This is very complicated transcendental equation for E, depending on $\tau$ parameter. There are two values of $\tau$, when this equation can be solved analytically:

i) $\tau = 0$. In this case we have only standard levels, which can be found from the poles of $\Gamma(1/2 - \lambda + P)$

$$1/2 - \lambda + P = -n_r; \quad n_r = 0,1,2... \quad (92)$$

ii) $\tau = \pm\infty$. In this case we have only additional levels, obtained from the poles of $\Gamma(1/2 - \lambda - P)$

$$1/2 - \lambda - P = -n_r; \quad n_r = 0,1,2... \quad (93)$$

Thus, in these cases one can obtain explicit expressions for standard and additional levels

$$E_{st,add} = -\frac{m\alpha^2}{2[1/2 + n_r \pm P]^2} = -\frac{m\alpha^2}{2\left[1/2 + n_r \pm \sqrt{(l+1/2)^2 - 2mV_0}\right]^2} \quad (94)$$

where signs (+) or (−) correspond to standard and additional levels, respectively.

We note that only the Eq. (92) was known till now. So the equation (91) and its consequences are new results.

Notice also that, in case $V_0 < 0$ we obtain well-known Kratzer potential [38], but now the condition (44) is not satisfied. Therefore there are no additional levels for Kratzer potential.

It is remarkable that the function (88) may be rewritten in unified form by using the following relation for the Whittaker functions [39]

$$W_{a,b}(x) = e^{-\frac{1}{2}x} x^{\frac{1}{2}+b} \frac{\pi}{\sin \pi(1+2b)} \left[ \frac{F(1/2+b-a, 1+2b; x)}{\Gamma(1/2-a-b)\Gamma(1+2b)} - x^{-2b} \frac{F(1/2-a-b, 1-2b; x)}{\Gamma(1/2+b-a)\Gamma(1-2b)} \right] \quad (95)$$

Then from (83), (88), (90) and (95) we derive

$$R(r) = C_1 \Gamma(1+2P) \Gamma(1/2-P-\lambda) \frac{\sin \pi(1+2P)}{\pi r} W_{\lambda,P}\left(\sqrt{-8mE}\, r\right) \quad (96)$$

Because the Whittaker function $W_{a,b}(x)$ has an exponential damping [39]

$$W_{a,b}(x) \underset{x \to \infty}{\approx} e^{-\frac{1}{2}x} x^a, \quad (97)$$

(97) corresponds to a bound state wave function which satisfies to the fundamental condition (43) for $0 < P < 1/2$ interval.

Therefore, for $\tau = 0, \pm\infty$ the standard and additional levels are obtained from (94) with corresponding wave functions

$$R_{st} = C_1 \rho^{-1/2+P} e^{-\frac{\rho}{2}} F(1/2+P-\lambda, 1+2P; \rho) \quad (98)$$

$$R_{add} = C_2 \rho^{-1/2-P} e^{-\frac{\rho}{2}} F(1/2-P-\lambda, 1-2P; \rho) \quad (99)$$

For arbitrary $\tau \neq 0, \pm\infty$ the energy can be obtained from the transcendental equation (91), while the wave function is given by (96).

The unified form (96) is also new result and it is a consequence of the SAE procedure.

According to [39] our function (96) takes the following form

$$R(r) = C_1 \Gamma(1+2P) \Gamma(1/2-P-\lambda) \frac{\sin \pi(1+2P)}{\pi \rho} e^{-\frac{\rho}{2}} \rho^{\frac{1}{2}-P} \Psi\left(\frac{1}{2}-\lambda-P, 1-2P; \rho\right) \quad (100)$$

where $\Psi(a,b,x)$ is one of the above mentioned solutions, (86), namely $y_5$. Its zeros are well-studied [39]: For real $a, b$ (note that in our case $a = \frac{1}{2} - \lambda - P$; $b = 1 - 2P$ are real numbers) this function has finite numbers of positive roots. However, for the ground state there are three cases where this function has no zeros:
1) $a > 0$; 2) $a - b + 1 > 0$; 3) $-1 < a < 0$ and $0 < b < 1$. Only the last case is interesting for us, because $a = \frac{1}{2} - \lambda - P$; $b = 1 - 2P$ and P is in the interval (47). It means

$$-1 < 1/2 - P - \frac{2m\alpha}{\sqrt{-8mE}} < 0 \quad (101)$$

In other words, the ground state energy, which is given by transcendental equation (91), must obey this inequality.

The wave function in form of (100) is also new.

In monograph [38] energy levels for alkaline metal atoms are written in Ballmer's form

$$E_{n'} = -R \frac{1}{n'^2} \quad (102)$$

where R is a Rydberg constant and $n'$ is the effective principal quantum number

$$n' = n_r + l' + 1 \quad (n_r = 0, 1, 2...) \quad (103)$$

while

$$l' = -1/2 \pm P = -1/2 \pm \sqrt{(l+1/2)^2 - 2mV_0} \qquad (104)$$

Only (+) sign was considered in front of the square root until now. In [38] $V_0$ was considered to be small and after expansion of this root, approximate expression for the standard levels was derived

$$E_{st} = -R \frac{1}{(n+\Delta_l)^2}; \quad n = n_r + l + 1 \qquad (105)$$

where

$$\Delta_l \equiv \Delta_l^{st} = -\frac{2mV_0}{2l+1} \qquad (106)$$

is so - called Rydberg correction (quantum defect) [23,38].

As regards of additional levels, this procedure is invalid, because $V_0$ is bounded from below according to (44). Approximate expansion for additional levels is possible only for $l = 0$. We have in this case

$$P = \sqrt{\frac{1}{4} - 2mV_0} \approx \frac{1}{2}(1 - 4mV_0) \qquad (107)$$

$V_0$ may be arbitrarily small, but different from zero, because in this case $P = 1/2$ and we have no additional levels.

One can easily obtain the existence condition of additional levels from (105) and (44) in diverse form

$$l < \Delta_l < l+1 \qquad (108)$$

If we use data of monograph [38], we obtain that for $l = 0$ states only Li, for $l = 1$ only K and for $l = 2$ only Cs satisfy (108) (i.e. they have additional solutions and it is necessary to carry out SAE procedure), and Na and Rb have no additional levels. The condition (108) helps us to determine which alkaline metals need SAE extension of Hamiltonian.

### 5B) The Klein-Gordon equation

Let us consider the Klein-Gordon equation in a central potential

$$(-\Delta + m^2)\psi(\mathbf{r}) = [E - V(r)]^2 \psi(\mathbf{r}) \qquad (109)$$

After the separation of angles, we derive the radial form of this equation

$$\left[ -\frac{d^2}{dr^2} - \frac{2}{r}\frac{d}{dr} + \frac{l(l+1)}{r^2} + m^2 - (E-V)^2 \right] R(r) = 0 \qquad (110)$$

and for the function $u = rR$, taking into account a condition (25), we have

$$u'' + \left[ (E-V)^2 - m^2 - \frac{l(l+1)}{r^2} \right] u = 0 \qquad (111)$$

It seems that even the Coulomb potential is singular by this equation. Now the following classification must be accounted for this equation

$$\lim_{r \to 0} rV(r) = 0 \quad \text{-} \quad \text{Regular} \qquad (112)$$

$$\lim_{r \to 0} rV(r) = -V_0 = const \quad \text{- Singular} \qquad (113)$$

i.e. the area of application of Eq. (111) becomes narrower. It is applicable only for potentials, satisfying to (112). Therefore the equation (111) may be used for potentials, which have less singularity than the Coulomb one, whereas in using of Eq. (110) no troubles appear.

**5C. "Hydrino" states in the Klein-Gordon equation with Coulomb potential**

We note that the problems of additional levels were discussed by other authors as well [39-42]. In particular, in [40] the Klein-Gordon equation as considered with $V = -\dfrac{\alpha}{r}$ Coulomb potential

$$R'' + \frac{2}{r}R' + \left[E^2 - m^2 - \frac{l(l+1)}{r^2} + \frac{2E\alpha}{r} + \frac{\alpha^2}{r^2}\right]R = 0 \tag{114}$$

The author underlines, that there must be levels below the standard levels (called, "hydrino" eigenstates), but he/she did not perform the SAE procedure.

Let consider this problem in more detail. First of all note that the equation (114) coincides with Eq. (82), but now

$$\rho = 2\sqrt{m^2 - E^2}; \quad \lambda = \frac{E\alpha}{\sqrt{m^2 - E^2}}; \quad P = \sqrt{(l+1/2)^2 - \alpha^2} > 0 \tag{115}$$

We must require $m^2 > E^2$ for bound states. Therefore one can use all the previous relations from valence electron model taking into account the definitions (115). In particular the SAE parameter now is

$$\tau = \frac{C_2}{C_1} \frac{1}{\left(2\sqrt{m^2 - E^2}\right)^P} \tag{116}$$

and for eigenstates we have the following equation

$$\frac{\Gamma(1/2 - \lambda - P)}{\Gamma(1/2 - \lambda + P)} = -\tau \left(2\sqrt{m^2 - E^2}\right)^P \frac{\Gamma(1 - 2P)}{\Gamma(1 + 2P)} \tag{117}$$

This is a new form, that follows by SAE procedure in the Klein-Gordon equation. For the edge points we derive the standard and additional levels in analogy with (117)

$$E_{st} = \frac{m}{\sqrt{1 + \dfrac{\alpha^2}{(1/2 + n_r + P)^2}}}; \quad n_r = 0,1,2... \tag{118}$$

$$E_{add} = \frac{m}{\sqrt{1 + \dfrac{\alpha^2}{(1/2 + n_r - P)^2}}}; \quad n_r = 0,1,2... \tag{119}$$

Exactly these (119) levels are called as "hydrino" levels in [39-42]. It is evident that the hydrino levels are analogical to $E_{add}$ states Eq.(94), but these two cases differ from each others. Particularly, it is possible to pass the limit $V_0 \to 0$ in the equation (82) and obtain Hydrogen problem. Usually this limiting procedure is used in traditional textbooks to choose between two signs in (94), while in (114) coupling constants for both terms in potential terms are mutually proportional ($\alpha$ and $\alpha^2$), and vanishing of one of them causes vanishing of another, so we turn to the free particle problem instead of Coulomb one. Moreover, as we mensioned above, in those papers [39-42] the SAE procedure was not used. They considered only two signs in front of square root in equation analogous to (94) and only (118) and (119) levels are considered, which correspond only to cases $\tau = 0$ and $\tau = \pm\infty$. Contrary to that case we performed SAE procedure, derived the Eq.(117) and take attention to the hydrino (when $\tau = \pm\infty$) problem.

The difference between standard and hydrino states manifests clearly in the nonrelativistic limit when $\alpha \to 0$, which must be performed by definite caution. The hydrino existence condition for such states folows from earlier constraints and the restriction $0 < P < 1/2$ It has a form

$$l(l+1) < \alpha^2 \qquad (120)$$

and it is evident that for states with $l \neq 0$ in transition to the nonrelativistic $\alpha \to 0$ limit the additional (hydrino) states disappear. Therefore we must consider only $l = 0$ states.

For the ground states ($n_r = l = 0$) we have

$$E_{st}^{(0)} = \frac{m}{\sqrt{2}} \sqrt{1 + \sqrt{1 - 4\alpha^2}} \qquad (121)$$

$$E_{Hyd} \equiv E_{add}^{(0)} = \frac{m}{\sqrt{2}} \sqrt{1 - \sqrt{1 - 4\alpha^2}} \qquad (122)$$

Expansion in powers of $\alpha$ gives

$$E_{st}^{(0)} = m\left(1 - \frac{\alpha^2}{2} - \frac{\alpha^4}{8}\right) \qquad (123)$$

$$E_{HYD}^{(0)} = m(\alpha + \alpha^3/2) \qquad (124)$$

It follows that the hydrino is very tightly bound system and sensitive to the sign of $\alpha$.

If we expand $l = 0; n_r \neq 0$ states till to order of $\alpha^2$, we derive

$$E_{st}^{(0)} = m\left(1 - \frac{\alpha^2}{2(n_r + 1)^2}\right) \qquad (125)$$

$$E_{HYD}^{(0)} = m\left(1 - \frac{\alpha^2}{2(n_r)^2}\right) \qquad (126)$$

Comparison of these two expressions shows that there appears some kind of degeneracy between the levels with $n_r + 1$ nodes of hydrino and energies for $n_r$ nodes of standard states. This degeneracy disappears in the next order.

The fact that the additional (hydrino[39-42] or peculiar [43,44] ) states of the $(n_r + 1)$ th $^1S_0$ state is nearly degenerate with the usual $n$ th $^1S_0$ state may facilitate a tunneling transition. Our description by the unified function analogous of (96), as a result of SAE procedure, gives a possibility of interpolation between them.

**5D) The Yukawa potential**

As a last application of Eq. (21) let us consider the Yukawa potential. According to common viewpoint (see, e.g. [8], Ch.28) the Yukawa potential is a spherically symmetric solution of the Helmholz wave equation

$$\nabla^2 \varphi - \mu^2 \varphi = 0 \qquad (127)$$

If we do not take attention to the appearance of the delta function, we would have a radial equation like [8]

$$\frac{1}{r} \frac{d^2}{dr^2}(r\varphi) - \mu^2 \varphi = 0 \qquad (128)$$

the solution of which is $r\varphi = Ce^{\pm\mu r}$ and in case of decaying asymptotic at infinity the Yukawa potential follows

$$\varphi = C\frac{e^{-\mu r}}{r} \tag{129}$$

However the application of the correct relation (21) gives

$$\nabla^2 \frac{e^{-\mu r}}{r} = \mu^2 \frac{e^{-\mu r}}{r} - 4\pi\delta^{(3)}(\mathbf{r})e^{-\mu r} \tag{130}$$

Interesting enough we found this equation in the earlier book [45]. It follows that the Yukawa potential is not a solution everywhere, but only outside the origin of coordinates. The Yukawa potential is the solution of the Helmholz wave equation with a source term on the RHS:

$$\nabla^2\varphi - \mu^2\varphi = -4\pi C\delta^{(3)}(r) \tag{131}$$

It was mentioned incorrectly in [7] that there is no need in imposing the boundary condition $u(0) = 0$ and it is sufficient to require regularity of solutions of the full radial equation. But it seems that in this particular case when the substitution (6) is applied, this requirement is equivalent to our restriction (25).

It is worthwhile to emphasize one important notion: Of course, to make the substitution (6) is not necessary at all. One can use other substitutions in course of solution of Eq. (127). In this case the conclusion of [7] becomes more transparent and lead to a new unexpected result.

Let discuss this viewpoint in case of Yukawa potential, i.e. of the Helmholz equation (127), rewriting it for the spherically symmetric solution as

$$\left[\frac{d^2}{dr^2} + \frac{2}{r}\frac{d}{dr}\right]\varphi - \mu^2\varphi = 0 \tag{132}$$

and instead of (6) make following substitution

$$\varphi = \frac{\chi(r)}{\sqrt{r}} \tag{133}$$

Denoting $z = \mu r$ we obtain an equation

$$\frac{d^2\chi}{dz^2} + \frac{1}{z}\frac{d\chi}{dz} - \left(1 + \frac{1}{4z^2}\right)\chi = 0 \tag{134}$$

The general solution of it is expressed in terms of modified Bessel functions [31]

$$\chi(z) = aI_{1/2}(z) + bK_{1/2}(z); \qquad z > 0 \tag{135}$$

Let us remember the asymptotic behavior for large and small arguments discussed above

$$I_P(z)_{z\to 0} \approx \left(\frac{z}{2}\right)^P \frac{1}{\Gamma(P+1)}, \qquad I_P(z)_{z\to\infty} \approx \frac{e^z}{\sqrt{2\pi z}}$$

$$K_P(z)_{z\to 0} \approx \left(\frac{z}{2}\right)^{-P} \frac{\Gamma(P)}{2}, \qquad K_P(z)_{z\to\infty} \approx \sqrt{\frac{\pi}{2z}}e^{-z} \tag{136}$$

We conclude that the second solution must be chosen owing the falling behavior at large values of argument. Therefore the solution of Eq. (132) is

$$\varphi = br^{-1/2}K_{1/2}(\mu r) \tag{137}$$

But [31]

$$K_{1/2}(z) = \sqrt{\frac{\pi}{2z}}e^{-z} \tag{138}$$

or

$$\varphi(r) = c \frac{e^{-\mu r}}{r} \qquad (139)$$

i.e. the Yukawa potential again.

However, as we saw above, unfortunately this is not the solution *everywhere*, because of singularity at the denominator.

It seems that this fact has very far reaching consequences. Namely, it turn out that the second function $K_P(z)$ is not a solution of the Bessel equation in spite of a widespread belief. Actually a straightforward transition of one-dimensional results of mathematical physics (theory of special functions, where the Laplacian is present) does not give necessarily the same things in three or more dimensions.

## 5. CONCLUSIONS

We have found a singularity like the Dirac delta function in process of reduction the Laplace equation in spherical polar coordinates, which was not mentioned earlier. The cornerstone in our consideration was a requirement of Dirac that the solution of the radial equation at the same time must be a solution of the full 3-dimensional equation.

On the basis of this observation we have proved that for removing this extra term from the radial equation it is necessary and sufficient to impose the reduced radial wave function by definite restriction, which has a form of the boundary condition at the origin, eq. (25). Moreover this condition is independent of whether the potential in the Schrodinger equation is regular or singular. The singular potential influences only the character of turning to zero of the radial function at the origin.

As regards of the full radial function $R(r)$, its equation is compatible with the primary (3-dimensional) equation (1) if the restriction (43) is satisfied. Therefore, to avoid the misunderstandings, it is preferable to work by the equation (4) in nonrelativistic and (110) in relativistic (Klein-Gordon equation) cases, correspondingly. Moreover only nonsingular solutions of full radial equation must be taken into account, only they are compatible with the full 3-dimensional equations.

The substitution (6) is convenient because the problem reduces to one dimensional one on the semi-axis. The real picture is as follows:

Particle in principle is able to move on the whole axis, but the effective potential is infinite for all negative values of argument. In this case the wave function is identically zero on the whole negative axis. The condition $u(0) = 0$ guarantees continuity of the wave function at $r = 0$. This provides the compatibility with the full equation and the equivalence to one-dimensional problem [14].

Above described situation takes place in spaces with dimensions three and more. Therefore in all equations of mathematical physics, where the Laplacian is involved, after the separation of angular variables the singular solutions, generally speaking, would not be the solutions of the primary equations.

If we shut eyes to a term with the delta function and formally use the reduced radial equation, then all results derived till now with the aid of this equation for regular potentials with regular boundary condition at the origin, remain valid. It is not an insignificant result from practical point of view.

However, when one considers singular potentials the use of equation for the full radial function $R(r)$ in parallel with the SAE procedure of the full radial Hamiltonian is necessary,. The appropriate examples, considered above, elucidate this statement.

# ACKNOWLEDGEMENTS

The Authors are grateful to professors M.A. Mestvirishvili and A.N. Kvinikhidze for valuable discussions and critical remarks. Moreover we thank the National Rustaveli Foundation (Grants D1/13/02 and FR/11/24) for the financial support.